# SCADS: Scale-Independent Storage for Social Computing Applications

Michael Armbrust, Armando Fox, David A. Patterson, Nick Lanham,
Beth Trushkowsky, Jesse Trutna, and Haruki Oh

University of California, Berkeley

{marmbrus, fox, pattrsn, nickl, trush, jtrutna, hoh}@cs.berkeley.edu

## ABSTRACT

Collaborative web applications such as Facebook, Flickr and Yelp present new challenges for storing and querying large amounts of data. As users and developers are focused more on performance than single copy consistency or the ability to perform ad-hoc queries, there exists an opportunity for a highly-scalable system tailored specifically for relaxed consistency and pre-computed queries. The Web 2.0 development model demands the ability to both rapidly deploy new features and automatically scale with the number of users. There have been many successful distributed key-value stores, but so far none provide as rich a query language as SQL. We propose a new architecture, SCADS, that allows the developer to declaratively state application specific consistency requirements, takes advantage of utility computing to provide cost effective scale-up and scale-down, and will use machine learning models to introspectively anticipate performance problems and predict the resource requirements of new queries before execution.

## 1. INTRODUCTION

Every popular website built on top of a traditional database eventually experiences problems with scaling the storage backend [9, 17, 18, 19, 20, 21]. Adopting increasingly powerful hardware solutions eventually becomes an insufficient strategy for maintaining a high request rate with low latency, leading operators to develop complicated systems built on top of many small clusters of relational databases. One example is the popular social networking website Facebook, which has over two billion dynamically generated page views per day [1]. Traffic of this magnitude results in over 23,000 page views a second, each of which could result in many queries to the database. Facebook's architects have been forced to respond to this load by federating their 1,800+ database instances into many independent, geographically-distributed clusters. In spite of this, the high request rate still necessitates supplementing their database infrastructure with over 25 terabytes of DRAM-based software caches [26]. Under this architecture, reads and writes are often routed to different datacenters, and coordinating the management of the partitioned data as well as the handling of cache lookups and invalidations required them to develop their own complex, proprietary system [24]. Notably, the system provides only eventual consistency, wherein a write to the system will not be seen by all users for a variable period of time. For Facebook and many other popular applications, this eventual consistency model has proven to be worth the performance advantages offered.

In the fast changing world of web development, time to market is often critical. Developers working on new websites generally do not have the time, resources, or expertise to implement a complex or scalable infrastructure at launch. As a result, most popular applications must undergo repeated architectural redesigns with the associated downtime, cost, and loss of momentum in the marketplace [4, 23]. As sites like these become more popular, re-architecting production systems "on the fly" becomes increasingly impractical.

The problem is further complicated by the fact that many Web 2.0 datasets are not easily partionable. Online communities create a user interaction graph that also functions as a means of site navigation. More links between users and their interests translates to a richer user experience. This phenomenon is an example of the network effect and could be summarized as: "The value of a site is derived from the interconnections between its users." Such interconnections between users are not amenable to database partitioning since every user may potentially interact with any other; i.e. one cannot rely on the presence of largely disjoint subsets in the graph. While shared-nothing architectures have been successfully employed to manage access to large, distributed data sets, their focus on providing full SQL over an optimally partitioned data-set is not well matched to these scenarios.

Another opportunity that could change the space of large-scale storage is the rise of utility computing. Prepackaged computing services like file storage and processing power can be rented by the hour, e.g. Amazon's Simple Storage System (S3) and Elastic Compute Cloud (EC2). The utility computing model allows developers to purchase only what they need at a very fine granularity (hours to minutes), with the opportunity to scale up and down as computing needs change. Therefore, we believe *rapid scale-down* is a new goal for massive storage systems, as there is now an economic benefit to doing so.

## 1.1 Data Scale Independence

Many of the problems described above are due to coupling scaling considerations — such as consistency and performance tradeoffs — too tightly to the application. Just as logical data independence allows you to change the conceptual schema without having to change the application, *data scale independence* allows the user base to grow by orders of magnitude without changing the application. Our architecture, Scalable Consistency Adjustable Data Storage (SCADS), provides data scale independence through three innovations:

**Performance-Safe Query Language** A scale-aware, introspective query language that is sufficiently general to enable efficient web programming, but provides strict scalability guarantees and predictable performance.

**Declarative Consistency-Performance Tradeoffs** A declarative language for reasoning about consistency and performance tradeoffs that allows developers to specify what correctness means for their application along with necessary performance SLAs, enabling the system to automatically tune itself to meet these parameters. Additionally, the system can provide guidance as to possible implementation costs based on machine learning performance models.

**Scaling Up and Down Using Machine Learning** The capability to use machine learning models to add and remove capacity to meet SLAs efficiently without downtime.

We explore the reasoning behind these innovations and describe our preliminary progress towards implementing them in a functioning system.

## 2. NEW REQUIREMENTS

We propose exploiting the special workload characteristics of Web 2.0 apps to address the previously described scaling problems with novel storage system designs. Any such design must make storage system scaling a first order priority, rather than expecting developers to apply and reapply an ad-hoc patchwork of temporary fixes.

## 2.1 Scaling Down As Well As Up

One of the most important considerations when designing a data storage system for this space is the ability to rapidly scale to handle more users. Figure 1 shows that one popular website, Animoto, went from using 50 machines to over 3400 machines in just three days [32].

We define "scaling" to mean *servicing more (or fewer) users while keeping the cost per user constant*. Due to the interactive nature of users' queries, the *response time for any given query must be invariant with respect to the number of users in the system*[1]. In order to support fast deployment, the system should be able to accommodate any number of additional users without the developers needing to modify any component of the storage system.

---

[1] Assuming the number of queries served by a web application is proportional to the number of active users, any query that performs a linear number of operations $w.r.t$ the number of users will result in polynomial load against the data storage system.

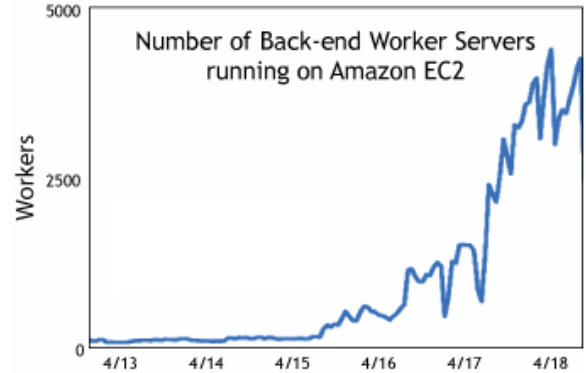

**Figure 1:** Animoto's viral growth caused them to go from tens of servers to 3400+ in only three days.

Utility computing — popular on many of these sites, including Animoto — changes the rules of scaling. With fine-grained billing done per machine hour, keeping idle servers active during non-peak times is a waste of money. In addition to the normal diurnal patterns, there are other events that cause huge spikes in traffic. For example, Facebook sees an increase in the number of photos posted the day after Halloween. Workloads such as this are particularly interesting, and difficult, because they involve a significant percentage of writes. Eventually being able to scale up to the number of active users is no longer sufficient, and a storage system must provide adaptive capabilities to both rapidly scale up for load spikes as well as to scale down to lower costs.

## 2.2 Declarative Consistency and Performance Specification

Eventual consistency is common in large scale storage systems, but can be confusing to developers and users. For example, Facebook users are confused when they do not see a post on someone's "wall" that they just made. In contrast, on Craigslist, the fact that a new listing will not appear in a search for five minutes is widely understood and considered acceptable by both developers and users.

Instead of these poorly-specified consistency models, we envision a simple declarative language that allows developers to easily reason about performance/consistency tradeoffs. Once the developer has declaratively specified what correctness means for a particular application in easy-to-understand terms like wall-clock time, the system should automatically use machine learning–based models of past performance to provision for future activity. Although other systems have presented tuning parameters like quorum requirements [7], we argue that is it more effective to allow the developer to state the correct behavior and let adaptive routines determine how to implement it.

## 2.3 Scale-Aware Query Language

Prior work has shown that large scale key-value storage can be a great tool for large-scale websites. Systems like Dynamo, BigTable, Cassandra, and Memcached can be found supporting almost every large website on the web. It is very difficult to program complex sites against such a limited interface, however, and even those systems which do have a SQL-like interface often lack key features, such as joins.

Conversely, a general-purpose language like SQL allows execution of slow queries and risks slowing down the rest of the system and upsetting users.

What developers really want is a system that can execute any query that is guaranteed to meet the "constant cost per user" scaling requirement, while also knowing ahead of time the consistency behavior and expected cost in terms of storage and processing to maintain the index, as determined by performance on past workload. This means that any update must be guaranteed to execute in less than $O(K)$ time, where $K$ is an application-specific constant. One example of this restriction is the limit of 5,000 friends per user on Facebook, allowing interesting joins to be done over all of a user's friends. However, a system like Twitter, where users can both "follow" and be "followed" by an unbounded number of users would not map into our system without modification.

## 3. SCADS ARCHITECTURE

In this section we describe how we meet the requirements presented in the previous section using the provisioning feedback loop shown in Figure 2, along with other techniques.

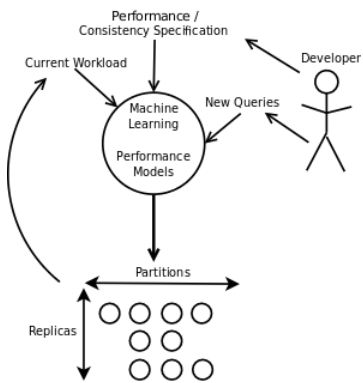

Figure 2: High level SCADS architecture. Each input and its effects on the system discussed below

### 3.1 Executing Queries

SCADS must execute thousands of queries per second with very low latency. To accomplish this, we impose the requirement that any query must be a lookup over a bounded contiguous range of an index[1]. While this precludes *ad hoc* queries, it guarantees that any single query requires at most one read from a small constant number of computers. The indices and views are automatically maintained according to the developer-specified consistency requirements using methods we now discuss.

### 3.2 Handling Updates

Due to the constraints discussed earlier, any update of data will most likely require updating several indices. To constrain the amount of work, we require that all queries be specified by the developer ahead of time in a restricted version of SQL that allows for cardinality constraints to be imposed. As developers list potential queries, SCADS analyzes them to ensure that they meet the scaling requirements discussed earlier. Specifically, we prohibit any structure for which any single update requires more than a constant amount of work.

---
[1]By "index" we mean any data structure containing pre-computed results. In particular, indices in SCADS may be more similar to materialized views than traditional database indices.

The given queries are compiled into a table of maintenance functions that are updated asynchronously and specify exactly what needs to be updated when data is changed. Update functions are required to run in $O(K)$ time, which is accomplished by limiting the number of lookup and update operations they can perform. This requirement, coupled with the fact that users can only do a bounded number of operations per day, guarantees we will meet our constant cost per user requirement while still supporting more interesting queries than can be performed over a simple index.

Figure 3 shows an example of such a table for a typical social network application. In this application, users are able to locate their friends, friends of their friends, and friends with upcoming birthdays. On an update or insert, SCADS will scan this table and call a specialized update function if the Table and Field match. The updates triggered by that function are prioritized according to the relaxed consistency model the developer has specified. Note that updatable structures may themselves be specified as tables, allowing an update on one to cause cascading updates to others.

| Index | Table | Field |
|---|---|---|
| friend_index | friendships | * |
| friends_of_friends_index | friend_index | * |
| birthday_index | profiles | birthday |
| birthday_index | friendship | * |

Figure 3: Table of typical index update operations for social network.

To find friends with upcoming birthdays a developer would submit a query template like this to the system:

```
SELECT profiles.* FROM friends f JOIN profiles p ...
WHERE f.f1 = <user_id> or f.f2 = <user_id> ...
ORDER by p.bday
```

The system would automatically construct an index where the first part of the key is a user_id and the second part is a birthday of one of that user's friends. This index entry would point to that friend's profile. The third and fourth rows of Figure 3 are the entries in the table that update this index. They specify that the index must be updated if a user changes her friend list or updates her birthday in her profile. While materializing these results will often result in a significant increase in the amount of storage needed, this is most likely not a problem as storage capacity is usually not a limiting factor for these applications.

This model restricts the types of queries SCADS can process. A query that is not a lookup in a pre-computed index will be rejected by SCADS, unlike in a traditional system which would allow the query to run slowly. In practice, we do not anticipate this restriction to become an obstacle for developers, as the queries needed by websites in this space are typically known in advance.

### 3.3 Implementation of Adjustable Consistency

We first define the axes that SCADS provides to developers for making performance consistency tradeoffs, and then discuss the methods we use for implementing this specification. Figure 4 summarizes the axes.

### 3.3.1 Performance-Consistency Specifications

| Axis | Effects | Example |
|---|---|---|
| Performance | Latency and Availability | 99.9% of requests succeed in <100ms |
| Write Consistency | Updates | Writes must be serializable, Last write wins |
| Read Consistency | Replication Latency | Stale data gone within 10 minutes |
| Session Guarantees | My Actions | I must read my own writes |
| Durability SLA | Data Durability | Data must persist with a 99.999% probability |

**Figure 4: The Axes of Consistency SCADS supports**

The first axis of consistency a developer can specify is a service level agreement (SLA) that states both the latency and availability requirements for accessing a specific set of data. These requirements take the form of "the 99.9th percentile request read latency must be under 100ms" and "99.99% of requests must succeed." Performance and failure models combined with current workload information will be used to automatically configure system parameters such as partitioning and replication. We believe machine learning algorithms will be able to make such predictions, based on previous successes using such algorithms to understand and predict performance in distributed systems [3, 6, 11, 33]. The results of these predictions can be shown to the user in the form of expected downtime vs. cost for implementing a policy to help them develop reasonable requirements.

Different applications require different handling of write conflicts, and our declarative consistency model provides a spectrum of consistency/performance tradeoffs for managing writes. For example, at one end of this spectrum a developer may state that writes to a given document type must be serializable, as in a traditional RDBMS. If conflicts are acceptable and can be intelligently resolved, the developer may specify a function that will merge conflicting writes. Finally, for applications when any write ordering would be correct, the developer could specify "last write wins" eventual consistency.

The possibility of reading stale data is inherent when using lazy replication. While developers and users understand this, they often want to know how long an update will take to propagate to all parts of the system. We allow developers to specify an upper bound on replication latency, e.g. "ten minutes." An example replication specification would be a statement defining the longest tolerable wall clock time between the time a write happens and the time all replicas have received a notification of it. Including such a specification would guarantee that no client would ever read data that is more stale than this upper bound time, the plausibility of which is evaluated by machine learning models. That is, if an update takes longer than the bound, a client query would stall until the updates can be confirmed. We later describe how this can be balanced with availability concerns.

Session guarantees as described by Terry et al. [31] provide users with a view of the database consistent with their own actions. We allow developers to specify whether sets of entities require read-your-own-writes or monotonic reads guarantees, the two most common cases required by web applications.

Distributed cluster-based systems already redefine the traditional notion of durability. While flushing a write to non-volatile storage was traditionally considered adequate, in large clusters nodes may appear and disappear without notification. As a result, durability may requiring persisting a write to multiple machines and possibly to non-volatile storage. Our declarative specification includes a durability SLA that allows developers to specify the probability that committed writes will actually persist given expected failure rates; for high volume but less-important data, such as old comments, relaxing this probability could save on replication costs.

Sometimes it is not physically possible to fulfill all of the specified requirements. There are often conditions in real world datacenters, such as network partitions or link congestion, that would prevent all requirements from being met simultaneously. In such cases, the system will using the developer-specified ordering of the requirements to decide which ones are more important. For instance, suppose we have an application that has specified 99.9% availability and read consistency with less than five minutes of lag. When two datacenters become disconnected, there would be contention between the availability SLA and the replication lag guarantees. If availability was prioritized over read consistency, then system would present stale data. However, if read consistency were prioritized over availability, then the system would return failure for read attempts. Failures of this type will be noted and used as input to the manager functions that re-provision the system in the future, either automatically or by notifying operators.

### 3.3.2 Consistency Implementation Methods

One very important difference between SCADS and previous relaxed consistency systems is the idea that the developer can specify wall-clock time bounds on "eventual" consistency. We have two main strategies for enforcing these guarantees: ordering of the relative importance of updates, and modeling of propagation time for future resource provisioning. Since the developers have specified propagation time bounds, the system knows the deadline for each update and can order them. The system will maintain a priority queue of updates, where the deadline for propagation is used as the priority. Not only does the priority queue allow the system to complete important updates first, but it allows us to easily detect when it is in danger of getting behind schedule. By using machine learning techniques to model the performance of these queues across the cluster in a utility computing environment, the system can proactively provision more machines before we risk violating the requirements of the developer. Similarly, in periods of less demand, the system can de-commit machines to lower costs while still maintaining SLAs.

## 3.4 Implementation Plan

We plan to leverage many existing technologies in our implementation of SCADS. First, Cassandra is an open-sourced structured storage system built on a P2P Network [16]. Starting with a proven scalable column-store as the underlying storage for SCADS, we will add asynchronous index updating, session guarantees, and automatic provisioning. To evaluate scalability we will use CloudStone [25], a Web 2.0 benchmark, running on thousands of instances on EC2. We plan to test ease of use by introducing SCADS to the undergraduate Ruby on Rails class at UC Berkeley [10]. If new Rails programmers with no experience building scalable systems can use SCADS to create a website that can support millions of synthetic users simulated via load generators, we will consider the project to have succeeded.

## 4. RELATED WORK

### 4.1 Distributed Databases

A great deal of academic and industry attention has focused on the design and construction of scalable, distributed databases. Teradata [30], Aster [2], and Greenplum [12] have demonstrated the capacity to scale to many nodes with near-linear performance enhancement on OLTP workloads. Based on the shared-nothing architecture [28], these systems scale by partitioning data into mostly disjoint subsets and distributing these subsets to independent machines. Data is partitioned and queries are processed in such a fashion so as to minimize inter-node data transfer [8]. As noted by Stonebraker [28], such systems work best when data is easily partitionable and access patterns are stable. Even newer systems like H-store [29] require a schema that can be converted into a tree of 1-n relationships, and do not promise interactive response for all queries. Social graphs, characteristic of Web 2.0 applications, are not amenable to such partitioning. Suffering from the "hairball problem", industry experience has shown that it is generally impossible to partition such graphs in a way that provides adequate performance over all queries [13].

Existing parallel databases don't focus on predicting the estimated running time or system-wide performance effects of a query submitted to the system. In a large Web 2.0 application, it is highly desirable to be able to estimate the effects of a query before running against a production database. In particular, it becomes crucial to identify and prevent the submission of queries whose running times are even linear with respect to the number of users.

Finally, many features and guarantees provided by distributed databases are stronger than required for social computing. In particular, modern web applications tolerate inconsistency to varying degrees and need not support *ad hoc* queries. SCADS exploits these two domain-specific restrictions to optimize performance.

### 4.2 Distributed Key-Value Stores

Many peer-to-peer (P2P) systems have attempted to address the problem of distributed data storage. Early systems like Gnutella were arbitrarily structured with no routing or replication schemes and relied on flooding to propagate queries for data throughout the system. Distributed hash table (DHT) systems like CAN [22] and Chord [27] have complex routing protocols to ensure that data can be located in a bounded number of hops. Early, unstructured systems are ill-suited to our environment due to their high latency. Even more structured systems can suffer from variable latencies as they route queries through multiple nodes, introducing unacceptably high delays for "unlucky" requests.

Pier [14] supported joins over a P2P overlay network like CAN or Chord, but does so over unconstrained amounts of data without using pre-computed indices. Initial results take seconds to return and full results can take minutes [14]; this delay would be unacceptable for our applications.

Cassandra [16], Dynamo [7], and BigTable [5] provide high performance, low latency data storage with replication. These systems provide durable, persistent data storage by replicating data in a distributed network. While each takes a different approach for storing and retrieving the data, the basic interface to the application is the same. Data is associated with a key (or a key and column name in the case of Cassandra and BigTable) on insert, and can be retrieved by that key. These newer key-value stores are designed for interactive applications and do not suffer from the multi-hop routing problem, but provide a limited data model. A limited data model inhibits programmers from rapidly developing and improving web applications and is undesirable in a competitive marketplace. In particular, the inability to perform arbitrary joins presents a significant roadblock to natural application development. By adding support for arbitrary indices to Cassandra, SCADS aims to maintain the advantageous performance characteristics of these systems while providing a richer data model. Specifically, SCADS will allow those joins capable of being safely executed in a production environment.

Megastore [15] is an effort by Google to add indices, ACID transactions, and schemas to BigTable. However, arbitrary joins are not natively supported. SCADS supports arbitrary joins for pre-computed queries with relaxed ACID semantics to ensure high performance.

## 5. CONCLUSION

We have described the key data management challenges faced by Web 2.0 applications: stringent response time and availability requirements, densely interconnected data, and exponentially increasing loads. We can improve performance in these environments by forsaking *ad hoc* queries, but developers still need a richer and more structured data model than key-value stores provide. Similarly, we can sacrifice single-copy consistency, but developers still need quantifiable and easy-to-understand consistency bounds. SCADS addresses the opportunity for a data management solution tailored to this class of applications. It provides scale independence by giving developers a rich, introspective query interface capable of anticipating performance problems, taking advantage of utility computing to support rapid scale-up and scale-down, and providing programmers with a declarative consistency-performance specification made possible with machine learning.

## 6. ACKNOWLEDGEMENTS

We would like to thank the following people for their invaluable feedback on the ideas in this paper: Henry Cook, Mike Franklin, Jeff Hammerbacher, and John Ousterhout.

This research is supported in part by gifts from Sun Microsystems, Google, Microsoft, Cisco Systems, Hewlett-Packard, IBM, Network Appliance, Oracle, Siemens AB, and VMWare, and by matching funds from the State of California's MICRO program (grants 06-152, 07-010, 06-148, 07-012,06-146, 07-009, 06-147, 07-013, 06-149, 06-150, and 07-008), the National Science Foundation (grant #CNS-0509559), and the University of California Indus-

try/University Cooperative Research Program (UC Discovery) grant COM07-10240.